%% file: manuscript.tex
\documentclass[10pt,twocolumn,letterpaper]{article}

\usepackage{iccv}
\usepackage{times}

\input{preprocessor}

\usepackage[pagebackref=true,breaklinks=true,letterpaper=true,colorlinks,bookmarks=false]{hyperref}

\usepackage[accsupp]{axessibility}  

\iccvfinalcopy 

\def\iccvPaperID{22} 

\ificcvfinal\fi

\begin{document}

\title{Thermal Image Processing via Physics-Inspired Deep Networks}

\author{Vishwanath Saragadam, Akshat Dave, Ashok Veeraraghavan, and Richard G.\ Baraniuk\\
Rice University, Houston TX\\
{\tt\small \{vishwanath.saragadam, akshat.dave, vashok, richb\}@rice.edu}
}

\maketitle
\ificcvfinal\thispagestyle{empty}\fi

\begin{abstract}
   \input{abstract.tex}
\end{abstract}

\section{Introduction}\label{section:intro}
\input{introduction.tex}

\section{Prior Work}\label{section:prior}
\input{prior.tex}

\section{Physics of Microbolometer Sensors}\label{section:physics}
\input{proposed.tex}

\section{Image Enhancement via Camera Motion}\label{section:multiframe}
\input{multiframe.tex}

\section{Regularizing Physics with Deep Networks}\label{section:nn}
\input{solving.tex}

\section{Experiments}\label{section:experiments}
\input{experiments.tex}

\section{Conclusion}\label{section:conclusion}
\input{conclusion.tex}

\section{Acknowledgements}\label{section:ack}
\input{ack.tex}

\begin{appendices}
	\section{Learning Details}\label{section:learning}
\input{sup_learning.tex}
	
	\section{Real Results}\label{section:sup_real}
	\input{sup_real.tex}
\end{appendices}

{\small
\bibliographystyle{unsrt}
\bibliography{refs}
}

\end{document}

%% file: preprocessor.tex
\pdfoutput=1
\usepackage{graphicx}
\usepackage{amsmath}
\usepackage[T1]{fontenc}

\usepackage{amssymb}
\usepackage{subcaption}
\usepackage{enumitem}
\usepackage{appendix}
\usepackage{animate}
\usepackage{comment}
\usepackage[numbers, sort&compress]{natbib}
\usepackage{url}
\usepackage{xcolor}
\usepackage{capt-of}
\usepackage{etoolbox}

\setlist{leftmargin=*}

\usepackage[ruled]{algorithm2e} 

\SetAlFnt{\small}
\SetAlCapFnt{\small}
\SetAlCapNameFnt{\small}
\SetAlCapHSkip{0pt}

\newcommand{\bfx}{{\bf x}}
\newcommand{\bfy}{{\bf y}}

\newcommand{\bfo}{{\bf o}}

\newcommand{\bfp}{{\bf p}}
\newcommand{\bfi}{{\bf i}}
\newcommand{\bfr}{{\bf r}}

\newcommand{\bfn}{{\bf n}}
\newcommand{\bfg}{{\bf g}}

\newcommand{\bpara}[1]{\vspace{0.5em}\noindent{\bf #1}}

%% file: abstract.tex
We introduce \textbf{DeepIR}, a new thermal image processing framework that combines physically accurate sensor modeling with deep network-based image representation.
Our key enabling observations are that the images captured by thermal sensors can be factored into slowly changing, scene-independent sensor non-uniformities (that can be accurately modeled using physics) and a scene-specific radiance flux (that is well-represented using a deep network-based regularizer).
DeepIR requires neither training data nor periodic ground-truth calibration with a known black body target--making it well suited for practical computer vision tasks.
We demonstrate the power of going DeepIR by developing new denoising and super-resolution algorithms that exploit multiple images of the scene captured with camera jitter.
Simulated and real data experiments demonstrate that DeepIR can perform high-quality non-uniformity correction with as few as three images, achieving a 10dB PSNR improvement over competing approaches.

%% file: introduction.tex
Long wave infrared (LWIR) thermal cameras capture a scene's intensity in the wavelengths spanning 8--14$\mu m$.
%
%
%
%
%
%
%
%
%
Thermal cameras in LWIR wavelengths find important applications in various scenarios including autonomous driving~\cite{fliradas2021}, robust computer vision~\cite{ivavsic2019human,krivsto2020thermal,hwang2015multispectral}, and large scale temperature monitoring~\cite{janssens2017deep}.
This democratization of thermal imaging is enabled by advances in low-cost uncooled microbolometer sensors.
%
\begin{figure}[!tt]
	\centering
	\includegraphics[width=\columnwidth]{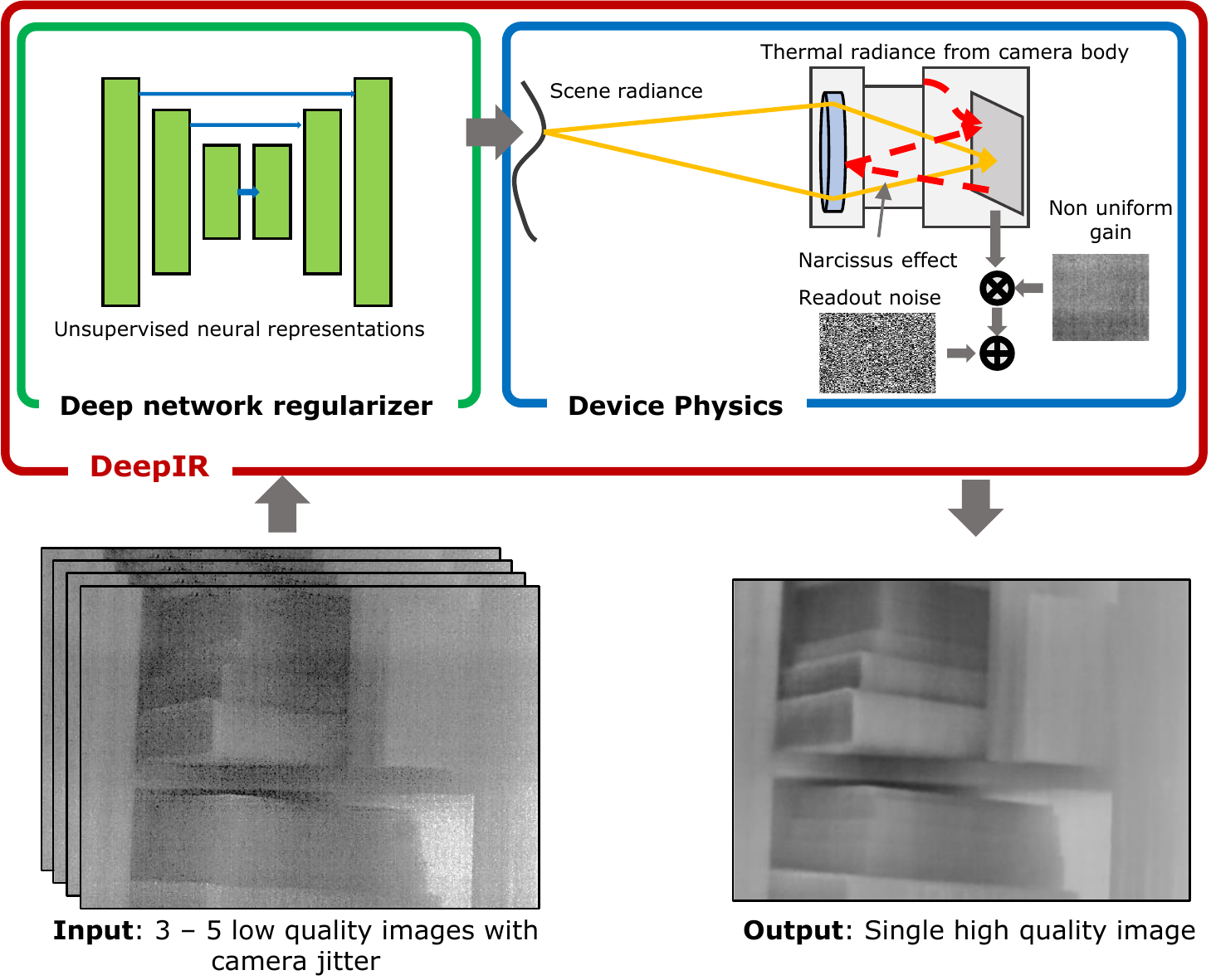}
	\caption{\textbf{DeepIR thermal image processing.} DeepIR is a novel thermal camera processing pipeline that combines physically accurate sensor modeling with deep networks to solve inverse problems in the thermal domain. We rely on capturing multiple, jittered images of the scene and then simultaneously estimates the scene's radiant flux by regularizing with a deep network-based regularization.}
	\label{fig:teaser}
\end{figure}
Despite the wide range of applications, uncooled microbolometer sensors face some unique challenges.
%
%
First, due to sensor-specific noise properties such as non-uniform per-pixel gain and high readout noise the signal to noise ratio is often low.
%
%
%
Second, the internal heating of the camera creates ``self-imaging,''  artifacts called the {\em narcissus effect}~\cite{vollmer2017infrared}.
It is hence imperative to augment the low-cost sensors with effective hardware and software solutions to produce high quality images.
\begin{figure*}[!tt]
	\centering
	\includegraphics[width=\textwidth]{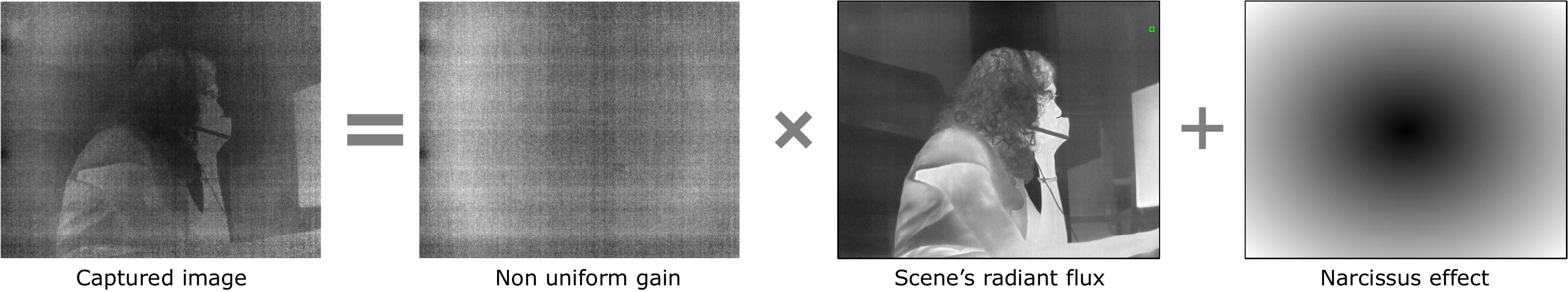}
	\caption{\textbf{Non-uniform noise in microbolometer thermal cameras.} This figure visualizes a simulation of image formation with a microbolometer sensor. Due to thermal changes within the camera, the final measurement suffers from spatially varying gain and offset.}
	\label{fig:nonuniformities}
\end{figure*}

There have been several approaches to enhance thermal images by combining multiple measurements~\cite{hardie2000scene,hardie2007map}, using data driven models~\cite{he2018single}, and multi modal fusion~\cite{rivadeneira2019thermal}.
%
%
Most approaches rely on strong assumptions about the spatial distribution of noise, such as the noise being unbiased or that the noise affects all pixels equally in a column.
In real images, such models do not completely capture the statistics of the noise, inevitably leading to poor recovery, or requiring dozens of images to produce high quality images.

%
%
%
%
%
%
A key observation about uncooled thermal sensors is that the image of a scene can be factored into a scene independent component and a scene dependent component (see Fig.~\ref{fig:nonuniformities}).
The scene independent component includes the gain and offset which arise due to slowly changing thermal conditions within the camera.
The scene dependent component includes the scene's radiant flux.
We exploit this observation by capturing multiple images of scene with camera motion which only affects the scene's radiant flux measurement and not the camera non-uniformities.
We then estimate the camera non-uniformities and the scene's radiant flux with a joint optimization approach.
To solve the inverse problem, we rely on the regularizing capabilities of convolutional neural network~\cite{ulyanov2018deep} which provides a concise representation for the scene's radiant flux.

The culmination of our efforts is a new image processing pipeline that we call {\em DeepIR} (pronounced ``deeper''), for Deep InfraRed image processing.
DeepIR can be used for recovering high quality images from a very small set of images captured with camera motion.
We demonstrate the advantages of DeepIR through several simulated and real experiments including non-uniformity correction, super resolution, and narcissus effect suppression.
	%
	%
%
%
An overview of the DeepIR pipeline is shown in Fig.~\ref{fig:teaser}.

The rest of the paper is organized as follows. We review the relevant prior work in section \ref{section:prior} and the physics of uncooled microbolometer sensors in \ref{section:physics}. This motivates our multi-frame measurement strategy explained in section \ref{section:multiframe}.
%
%
We then dive DeepIR into image enhancement with deep network-based representation in section \ref{section:nn}, and compare against prior art in \ref{section:experiments}.
%
%
We conclude in section \ref{section:conclusion} with some notes on future directions.
To enable further research in thermal image processing, we have made our source code and datasets publicly available\footnote{\url{https://github.com/vishwa91/DeepIR}}.

%% file: prior.tex
%
Thermal cameras are based either on photonic sensors or microbolometers.
Photonic sensors rely on semicondoctors to absorb light photons, whereas microbolometers utilize a temperature-dependent resistance to convert thermal radiation to digital output.
%
%
%
%
%
%
%
Due to low manufacturing costs and no external cooling, microbolometer cameras are cheap and compact -- making them amenable for several vision-based tasks.
%
We hence focus on microbolometer-based cameras throughout the paper.

Most low-cost microbolometer cameras do not employ thermal stabilization of the focal plane array (FPA), making the measurements highly sensitive to temperature changes.
This results in a slowly drifting non-uniformity that degrades the quality of the image (see Fig.~\ref{fig:nonuniformities}).
%
%
It is hence important to correct for the sensor-specific non-uniformities to obtain accurate measurements.
Methods for non-uniformity correction (NUC) for microbolometer sensors can be broadly categorized as hardware-based or software-based.

\bpara{Hardware approaches.} NUC can be performed reliably with an image of flat blackbody at a known temperature.
The most popular solution in this approach is the so-called shutter-based flat field (FFC) which relies on periodically capturing images with a closed shutter.
Such approaches are not ideal as the mechanical components induce vibrations, and significantly increase power consumption.
%
%
%
%
%
%
%
Solutions which involve a semi-transparent shutter have been proposed that remove the necessity to close the camera~\cite{olbrycht2015new} but require extremely careful calibration of output reference for each operating temperature.

\bpara{Software approaches.}~These exploit the unique properties of microbolometer to correct for non-uniformities, either using a single image~\cite{he2018single,liu2018shutterless} or multiple images~\cite{hardie2000scene,hardie2007map,wolf2016modeling}.
%
%
%
%
%
%
Of particular interest in this regard is the work by Hardie et al. \cite{hardie2000scene,hardie2007map} which models the image formation as a product of fixed camera-specific gain, and a moving scene-specific radiance.
Parameters are then estimated by solving a simple least squares problem.
%
%
%
While the approach is promising, the estimated image is sensitive to accuracy of registration and the initial estimate.

DeepIR is inspired by the works of Hardie et al.~\cite{hardie2000scene,hardie2007map} that combines multiple images of a scene captured with camera motion.
Our core contribution is an end-to-end pipeline that jointly estimates the camera non-uniformities, and the scene's radiant flux.
We achieve this by regularizing the inverse problem with a concise deep prior-based image representation.
%
%
%
%
%

%
%

%% file: proposed.tex
Our goal is to recover a high quality image for a few, low quality thermal images corrupted by non-uniform noise.
%
%
%
We first present a simple image formation model which motivates the DeepIR image processing pipeline.

\bpara{Sensor modeling.} Consider a single pixel in the 2D sensor. Let $\Phi_\text{scene}$ be the radiant flux incident on the pixel and $\Phi_\text{fpa}$ be flux emitted by the pixel. 
Let $C$ be the thermal capacitance of the microbolometer pixel, and $G$ its thermal conductance.
The resulting change in temperature $\Delta T$ is related to the above quantities by the energy conservation equation~\cite{vollmer2017infrared}
\begin{align}
	\alpha(\Phi_\text{scene} - \Phi_\text{fpa}) = C\frac{d\Delta T}{dt} + G\Delta T\label{eq:fpa}.
\end{align}
%
%
%
Unlike photonic sensors, a microbolometer pixel is \emph{always} exposed to the scene's radiant flux resulting in the so-called \emph{thermal inertia} that prevents abrupt temperature changes in the sensor.
Thermal inertia produces a characteristic motion blur with exponentially decaying point spread function that varies spatially~\cite{ramanagopal2020pixel,oswald2010motion}.
Assuming the incident flux changes from $\Phi_1$ to $\Phi_2$ in a step manner, we can model the change in temperature of the pixel as~\cite{vollmer2017infrared}
\begin{align}
	\Delta T = \frac{\alpha\Phi_1}{G}e^{-\frac{t}{\tau}} + \frac{\alpha\Phi_2}{G}\left(1-e^{-\frac{t}{\tau}}\right),
\end{align}
where $\alpha$ is the conversion efficiency of the microbolometer and $\tau = \frac{C}{G}$ is the time constant of the microbolometer and is a measure of thermal inertia of the pixel.
This change in temperature manifests as change in resistance of the microbolometer $\Delta R$
\begin{align}
	\Delta R = \beta R_\text{avg} \Delta T,
\end{align}
where $\beta$ is the temperature coefficient of the pixel, and $R_\text{avg}$ is the average resistance of the microbolometer over the measurement duration.
Assuming the pixel reaches steady state within the integration time,
\begin{align}
	\Delta R = \frac{\beta R_\text{avg} \alpha (\Phi_2 - \Phi_\text{fpa})}{G}.
\end{align}
If $I$ be the current flowing through the microbolometer
\begin{align}
	\Delta V &= I \Delta R = \frac{\beta R_\text{avg} \alpha (\Phi_2 - \Phi_\text{fpa})}{G}\nonumber\\
			 &= \frac{I\alpha\beta R_\text{avg}}{G} \Phi_2 - \frac{I\alpha \beta R_\text{avg}}{G}\Phi_\text{fpa}\nonumber\\
\implies \Delta V &\equiv g \Phi_2 + o\label{eq:phi2v},
\end{align}
where $g, o$ are slope and intercept respectively relating the input radiant flux to output voltage.
Extending the analysis to all pixels in the sensor for time instance $t$
\begin{align}
	\Delta V(u, v, t) &= g(u, v, t) \Phi_2 (u, v, t) + o(u, v, t).
\end{align}
Incorporating readout noise in the equation we obtain
\begin{align}
	y(u, v, t) &= g(u, v, t) x(u, v, t) + o(u, v, t) + n(u, v, t)\label{eq:camera},
\end{align}
where $u, v$ are camera pixel coordinates, $y$ is the digital output of the camera, and $x$ is scene's radiance that we wish to estimate.
%
%
Since microbolometers require a small bias current to operate, the temperature changes within the housing, leading to non-uniformities in gain and offset.

\bpara{Factors affecting offset.}
Prior works largely assumed that the offset terms in $o(u, v)$ are independent identically distributed Gaussian random  variables. 
However, practical systems have offset contributions from sources that are highly structured, such as internal heating of the camera's housing, or reflections off of the optical subsystem.
%
%
%
It is possible to correct for the offset by periodically capturing image of an external black body -- however this approach is not always feasible.
We instead model the offset term as a spatially smoothly varying signal which can be estimated along with gain and scene's radiant flux.
%
%
%
%

%% file: multiframe.tex
\begin{figure}[!tt]
	\centering
	\includegraphics[width=\columnwidth]{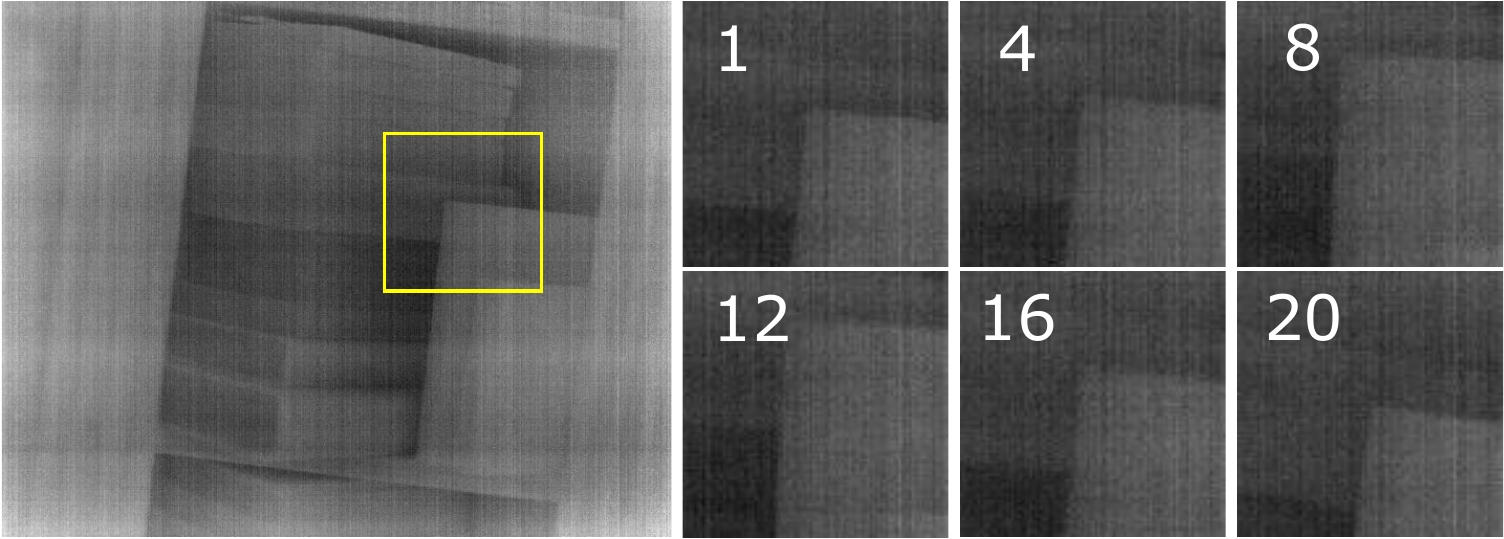}
	\caption{\textbf{Effect of camera jitter.} Camera motion affects only the radiant flux entering the camera and not the non-uniformities.}
	\label{fig:motion}
\end{figure}

Each frame captured by microbolometer camera is corrupted by an offset term and gain term that is specific to each camera and its operating temperature, which results in an ill-posed system of equations.
%
%
%
These gain and offset terms are intrinsic properties of the camera and change slowly over time.
This implies, if we were to capture multiple images of the scene over a short interval, any camera motion affects only the scene's radiant flux and not the non-uniformities due to the sensor.
This is visualized in Fig. \ref{fig:motion} where a sequence of twenty images was captured over a short period.
Evidently, the non-uniformities do not change over the duration of twenty frames.
\begin{figure}[!tt]
	\centering
	\includegraphics[width=\columnwidth]{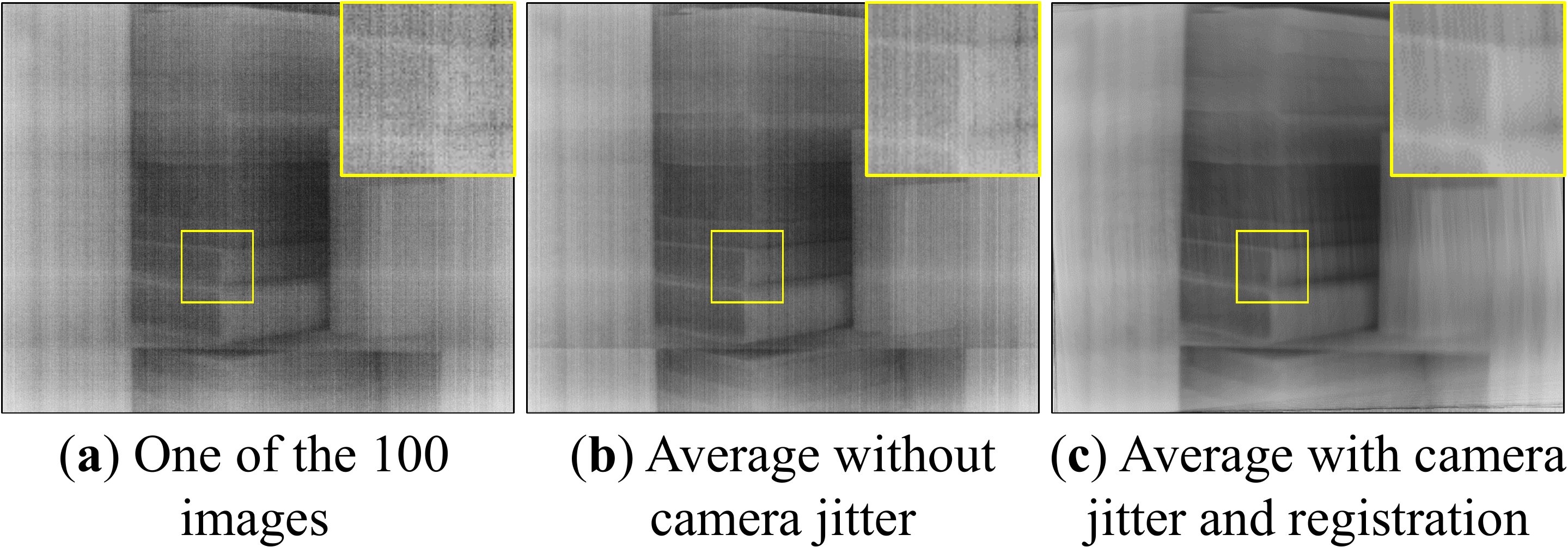}
	\caption{\textbf{Advantages of camera jitter.} Small camera motion, while undesirable in visible cameras, helps reduce the effect of slowly changing gain and offset in thermal images.}
	\label{fig:jitter}
\end{figure}
This inherent separation motivates our approach --- while jitter in the camera is undesirable in visible cameras, it is highly advantageous in the microbolometer camera.
As a simple example, consider a capture of $100$ images of a scene with and without camera jitter, shown in Fig.~\ref{fig:jitter}.
A simple averaging does not remove the non-uniformities in the absence of camera jitter.
However if we capture with camera jitter and then register all frames to the first reference frame and average, we obtain a relatively noise-free image.
DeepIR and averaging rely on camera jitter to recover the scene's radiant flux; however DeepIR requires far fewer images due to a combination of device physics and concise image representation.

\subsection{Modeling multiframe capture}
To regularize the inverse problem, we model the gain and offset terms to be constant for the duration of $L$ frames.
%
%
Further, we assume that every frame can be represented as a geometric transformation of the first frame, which includes rigid, affine, or perspective transforms.
The overall model
\begin{align}
	y(u, v, t_k) &= g(u, v) (x(f_k(u), h_k(u)) + o(u, v))\cdots\nonumber\\
	&\cdots + n(u, v, t_k),
\end{align}
where $f_k, h_k$ are functions relating pixels in $k^\text{th}$ frame to first frame.
Vectorizing all representations, we obtain
\begin{align}
	\bfy_k &= \bfg \odot \left(M_k \bfx_0 + \bfo\right) + \bfn\label{eq:modelvec},
\end{align}
where $\odot$ is element-wise multiplication, $M_k$ is the linear operator to perform the geometric transformation, $\mathbf{g}$ is the gain vector, $\bfx_0$ is the noise-free image, and $\bfo$ is the offset term.
Our goal is to recover the image $\bfx_0$.
For $L$ frames and $N$ pixels per frame, we have $N$ parameters each from the gain, offset, and latent images, and $8$ parameters from the geometric transformation assuming a generalized perspective transformation.
Overall, we have $LN$ equations and $3N + 8(L-1)$ unknowns.
\subsection{How much should we jitter the camera?}
The amount of jitter needed to accurately recover the scene's radiant flux is highly dependent on the nature of the non-uniformities.
Intuitively, the more correlated the spatial non-uniformities, the more the camera needs to jitter.
To understand the reason, consider the sequence of $L$ images, $Y_k(u, v) = G(u, v)X_0(u_k, v_k)$, where we assume that the offset is zero.
Let us assume that each image is registered back to the reference frame giving us
\begin{align}
	\widehat{Y}_k(u, v) &= G(\widehat{u}_k, \widehat{v}_k)X_0(u, v),
\end{align}
where $G(\widehat{u}_k, \widehat{v}_k)$ is the resultant gain after registering $X_k(u, v)$.
Then averaging the frame yeilds
\begin{align}
	\widehat{X}_0 &= \frac{1}{L}\sum_{k=1}^L \widehat{Y}_k(u, v)
				  = \frac{1}{L}X_0(u, v)\sum_{k=1}^L G(\widehat{u}_k, \widehat{v}_k).
\end{align}
The variance of estimate at pixel $(u, v)$ is then
\begin{align}
	&\sigma^2_{u, v} = \text{Var}\left(\frac{1}{L}X_0(u, v)\sum_{k=1}^L G(\widehat{u}_k, \widehat{v}_k)\right)\nonumber\\
				&= \frac{1}{L^2}X_0^2(u, v) \left(\sum_{k=1}^L \text{Var}(g_k) + 2\sum_{p=1}^L\sum_{q=1}^{L}\text{Cov}(g_p g_q) \right)\label{eq:var}.
\end{align}
Equation \eqref{eq:var} states that the variance of estimate depends on the autocorrelation function of the gain.
Assuming the autocorrelation function monotonically decreases with distance, it is intuitive to see why a more correlated gain requires larger jitter.

\begin{figure}[!tt]
	\centering
	\centering
	\includegraphics[width=\columnwidth]{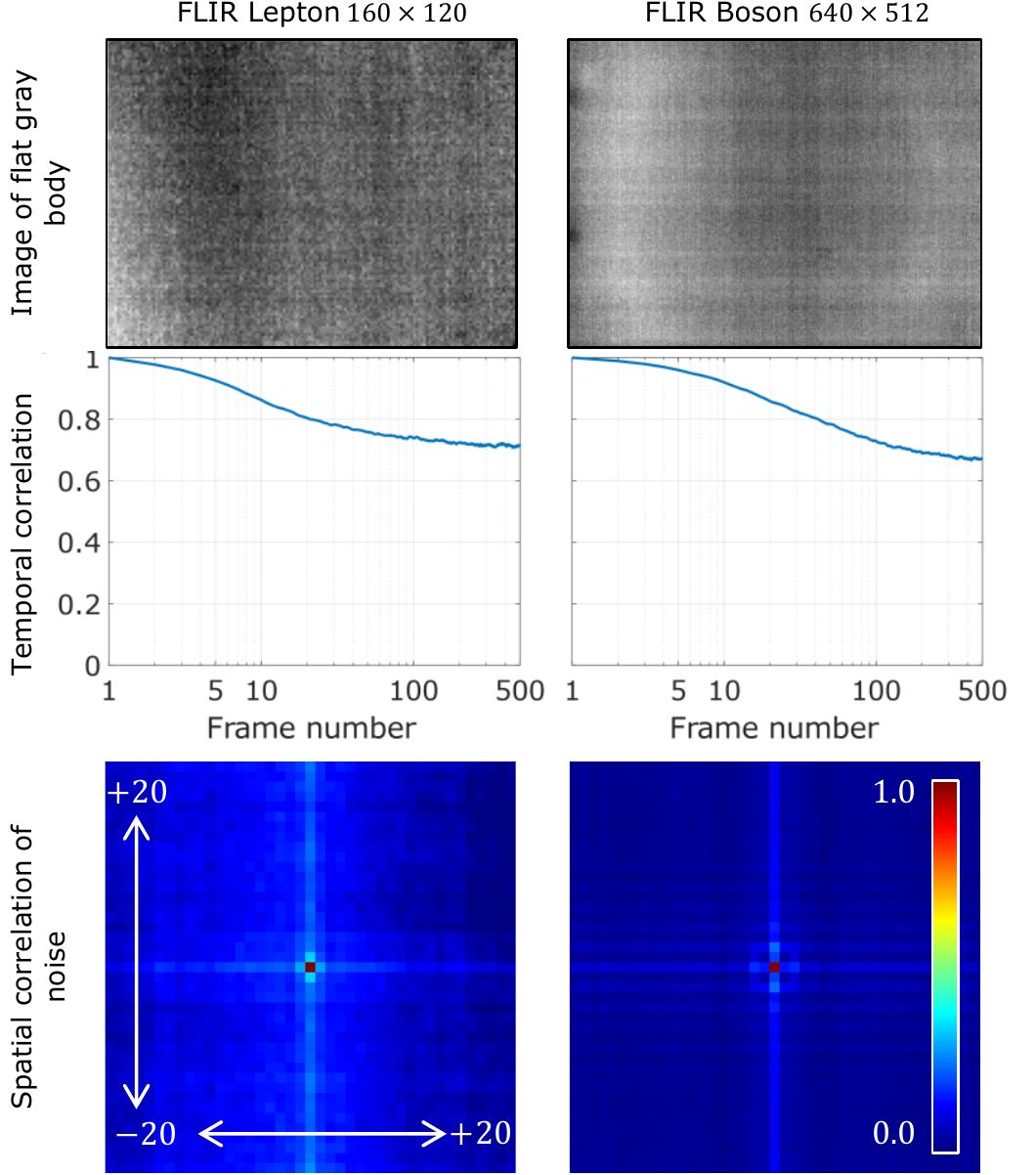}
	\caption{\textbf{Statistics of non-uniformities.} The non uniformities associated with thermal cameras have spatial and temporal correlations, which allows us to choose the minimum amount of jitter, as well as the maximum number of frames that are needed to obtain a high quality estimate of the scene's radiant flux.}
	\label{fig:stats}
\end{figure}

In practice, it is difficult to estimate the autocorrelation of the gain as it is a complex function of temperature of operation, and the electronic circuitry.
To obtain an empirical estimate of the amount of jitter needed, we imaged a flat black body with the low resolution FLIR lepton, and the medium resolution FLIR Boson cameras.
We then computed spatial autocorrelation by cropping random patches and computing cross correlation within a neighborhood of $50$ pixels on all sides.
Figure \ref{fig:stats} shows the captured image, and the temporal and spatial autocorrelation functions for both cameras.
We make three observations here.
First, the temporal autocorrelation gracefully reduces from $1.0$ to $0.6$ over $500$ frames, with value being greater than $0.8$ for up to $20$ frames.
This implies we can assume approximately constant non-uniformities for up to $20$ frames for both cameras.
Second, the spatial autocorrelation function is dominated along the horizontal and vertical axes --- this is expected since the microbolometer cameras are equipped with a rolling shutter readout circuitry~\cite{alhussein2016simulation}.
An immediate implication of this observation is that we cannot achieve noise reduction by just horizontal or vertical shifts, we need a combination of the two.
Third, the Lepton camera has an autocorrelation greater than $0.1$ over a shift of $3$ pixels, and the Boson camera over $5$ pixels on either sides.
Hence we require non-axial shifts of $3$, and $5$ pixels, respectively, for the two cameras to ensure high quality reconstruction with a small number of images.

%

%% file: solving.tex
In an ideal scenario, we can estimate the non-uniformities, motion parameters, and the scene's radiant flux from as few as $4$ frames.
However, due to both signal and readout noises, the inversion is often unstable.
Hardie et al.~\cite{hardie2000scene} approached this problem by assuming that the images to be registered, and posing \eqref{eq:modelvec} as a least squares problem with $20$ input frames.
This may not be feasible, as the camera temperature, or the scene may change within that duration.
%
%
Moreover, in the presence of severe noise, obtaining a reliable registration is difficult.
A simple extension is to then jointly optimize for the unknown registration
\begin{align}
	\min_{\bfg, \bfo, M_k, \bfx_0} \sum_{k=1}^{N} \| \bfx_k -  \bfg \odot M_k \bfx_0 - \bfo\|^2 + TV(\bfx_0)\label{eq:physics},
\end{align}
where $TV(\cdot)$ is the 2D total variation norm, acting as a regularizer.
However, the approach fails to converge in the absence of a good initial registration.
%

This inverse problem can be made tractable if we have a concise representation for the scene's radiant flux.
For images in the visible domain, there are several compelling ways of concisely representing images, including analytical signal models~\cite{chang2000adaptive}, or learned representations~\cite{aharon2006k,rick2017one}.
%
%
Learned models are better tailored to the statistics of real world images and hence have been the choice for image representation.
%
%
%
In the presence of a very large pool of data (including noisy and noise-free pairs), it is possible to learn good data-driven models for thermal imaging.
%
%
%
However, due to device-specific noise statistics of each microbolometer camera, such a data-driven approach may not be practical.
%

\subsection{Deep network as a regularizer}
%
Recent works by Ulyanov et. al.~\cite{ulyanov2018deep} on deep image prior have shown that the inductive bias of a convolutional neural networks act as concise priors for images.
Specifically, given a fixed input (commonly random noise) $\mathbf{n}$ to a neural network $\mathcal{N}$, deep image prior seeks to solve the following optimization problem,
\begin{align}
	\min_{\mathcal{N}} \| \mathbf{x} - \mathcal{N}(\mathbf{n})\|^2 + \lambda \mathcal{R}(\mathbf{x}),
\end{align}
where $\mathbf{x}$ is the signal of interest, and $\mathcal{R}$ is a regularizer specific to signal domain, such as the total variation norm for images.
We observe here that the weights of the network $\mathcal{N}$ are learned only with \emph{an} instance of the signal, and not on a pool of data.
Such a representation can then used to regularize a range of inverse problems including denoising, super resolution, and inpainting.

\subsection{Combining physics and neural representations}
Armed with our insights into the sensor physics and ability to concisely regularize images, we now explain how we can efficiently solve for the sensor and scene parameters.
We model the scene's radiant flux as the output of a neural network, specifically $\bfi_0 = \mathcal{N}(\bfp)$, where $\mathcal{N}$ is a convolutional neural network, and $\bfp$ is a fixed (possibly random) input.
We then solve the following optimization problem
\begin{align}
	\min_{\bfg, \bfo, M_k, \mathcal{N}} \| \bfx_k -  \bfg \odot M_k \mathcal{N}(\bfp) - \bfo\|^2
	\label{eq:dip}
\end{align}
This approach not only preserves the image formation, as well as sensor specific noise characteristics, but also incorporates a concise, non linear representation for image, that has been shown to produce promising results.
By optimizing the registration, gain, and the latent image simultaneously, DeepIR accurately estimates the parameters of the system and scene.
%
%
%
%
We make no further assumptions about the structure of the scene, or the non-uniform noise, implying that it works well for cameras with or without shutter-based NUC.
%
%
Since our approach is modular by separating out the device physics, and the choice of image representation, we can replace the deep image prior with other neural representations such as the deep decoder~\cite{heckel2018deep}, or implicit neural representations~\cite{sitzmann2020implicit,tancik2020fourier}.
We show all our experiments with the deep image prior, but extensions with other representations is straightforward.

\bpara{Solving linear inverse problems.}
DeepIR can be applied to several linear inverse problems of the form
\begin{align}
	\widehat{\bfr}_k = \Phi_k \bfr_k
	\implies \bfx_k = \bfg \odot (\Psi_k \bfr_k) + \bfo
\end{align}
where $\Psi_k$ is a linear operator.
This approach enables us to solve several problems including NUC, super resolution, deconvolution, and optical flow estimation.
As an exmaple, we showcase super resolution using DeepIR, where $\Psi_k$ is a downsampling operator.

%% file: experiments.tex
We now demonstrate the effectiveness of DeepIR over a diverse set of simulated and real experiments.
\subsection{Simulations}
We first focus on denoising images. To quantitatively compare various approaches, we took a clean thermal image and added fixed pattern noise, emulating effects due to the readout circuitry specific to microbolometer cameras~\cite{alhussein2016simulation}.
Each image was formed in the following manner,
\begin{align}
	Y_k = \mathcal{P}(GS_{\Delta x_k, \Delta y_k} (R_{\theta_k}(X_0))),
\end{align}
where $\mathcal{P}$ is poisson noise operator, $G$ is a random, columnwise gain, $S$ is the shift operator, and $R$ is the rotation operator.
We then evaluated the following approaches:
\begin{itemize}
	\item \textbf{Temporal averaging.} We compensate for camera motion by registering the $N$ images and then average them to obtain the latent image.
	\item \textbf{Hardie et al. \cite{hardie2007map}.} We solve the optimization problem in eq. \eqref{eq:physics} with TV prior on image.
	\item \textbf{He et al. \cite{he2018single}.} We used the deep learning-based approach proposed in \cite{he2018single}. For multi-frame comparison, we denoised each individual frame, and then registered and averaged them.
	\item \textbf{DeepIR.} We solve the optimization in \eqref{eq:dip}.
\end{itemize}
We obtained initial registration between images using the pyramidal registration scheme proposed in \cite{thevenaz1998pyramid}.
We used the same deep network architecture as proposed in \cite{ulyanov2018deep} for DeepIR.
%
%
For all optimization problems, we used PyTorch~\cite{NEURIPS2019_9015}.
Further details about training are provided in the supplementary document.

\begin{figure}[!tt]
	\centering
	\includegraphics[width=\columnwidth]{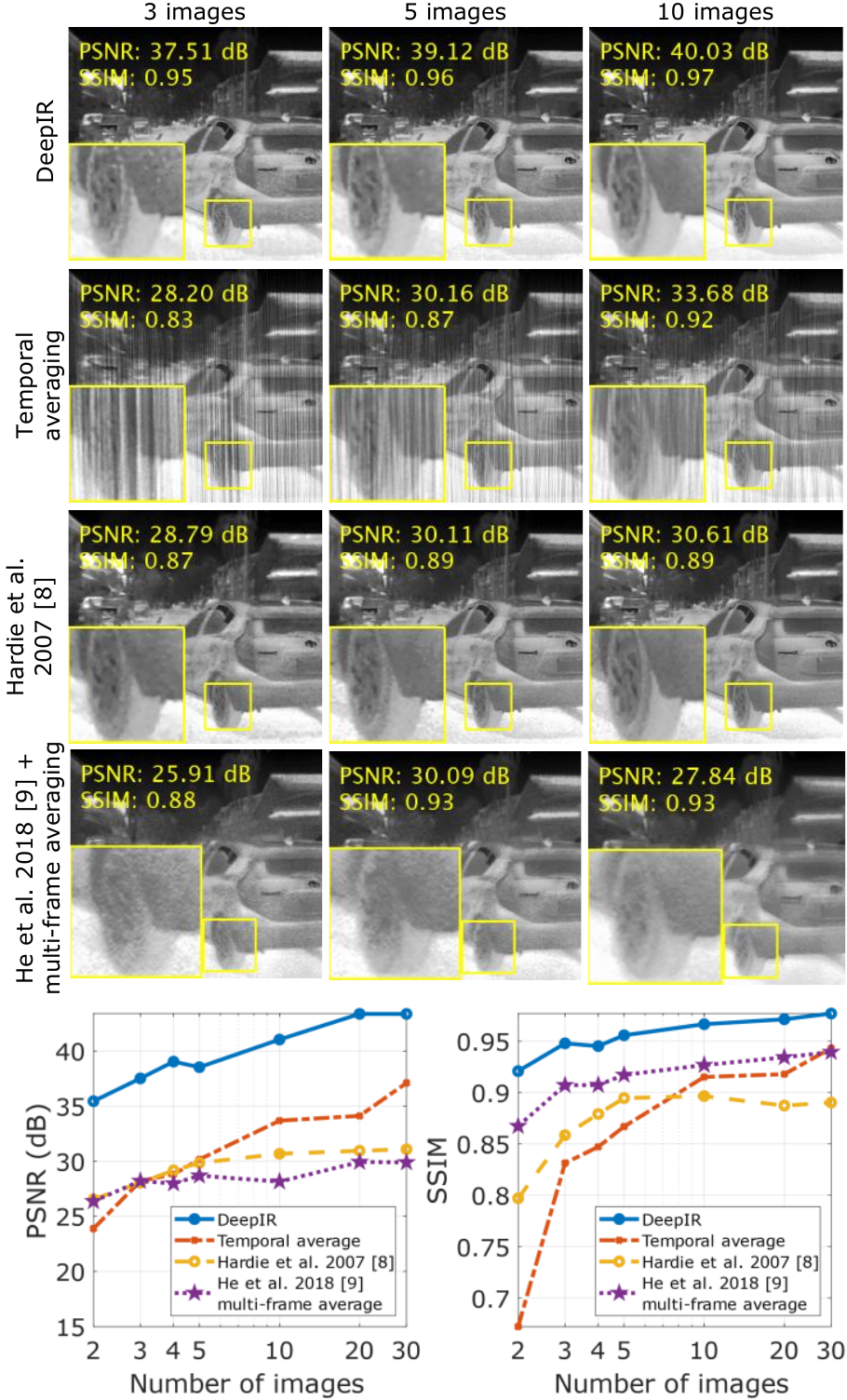}
	\caption{\textbf{Performance vs.\ number of images.} Across the board, DeepIR outperforms other techniques for denoising and NUC.}
	\label{fig:sim}
	\vspace{-1em}
\end{figure}

\bpara{Non-Uniformity Correction.} Figure \ref{fig:sim} shows a visualization of various approaches for varying number of images.
We observe that incorporating sensor model into the optimization problem enables a significant increase in accuracy, as can be seen in the results of \cite{hardie2007map}, and DeepIR.
The advantage of combining sensor model and deep networks is evident in the results of DeepIR, where the reconstruction is significantly better even with as few as three images.

\begin{figure}[!tt]
	\centering
	\includegraphics[width=\columnwidth]{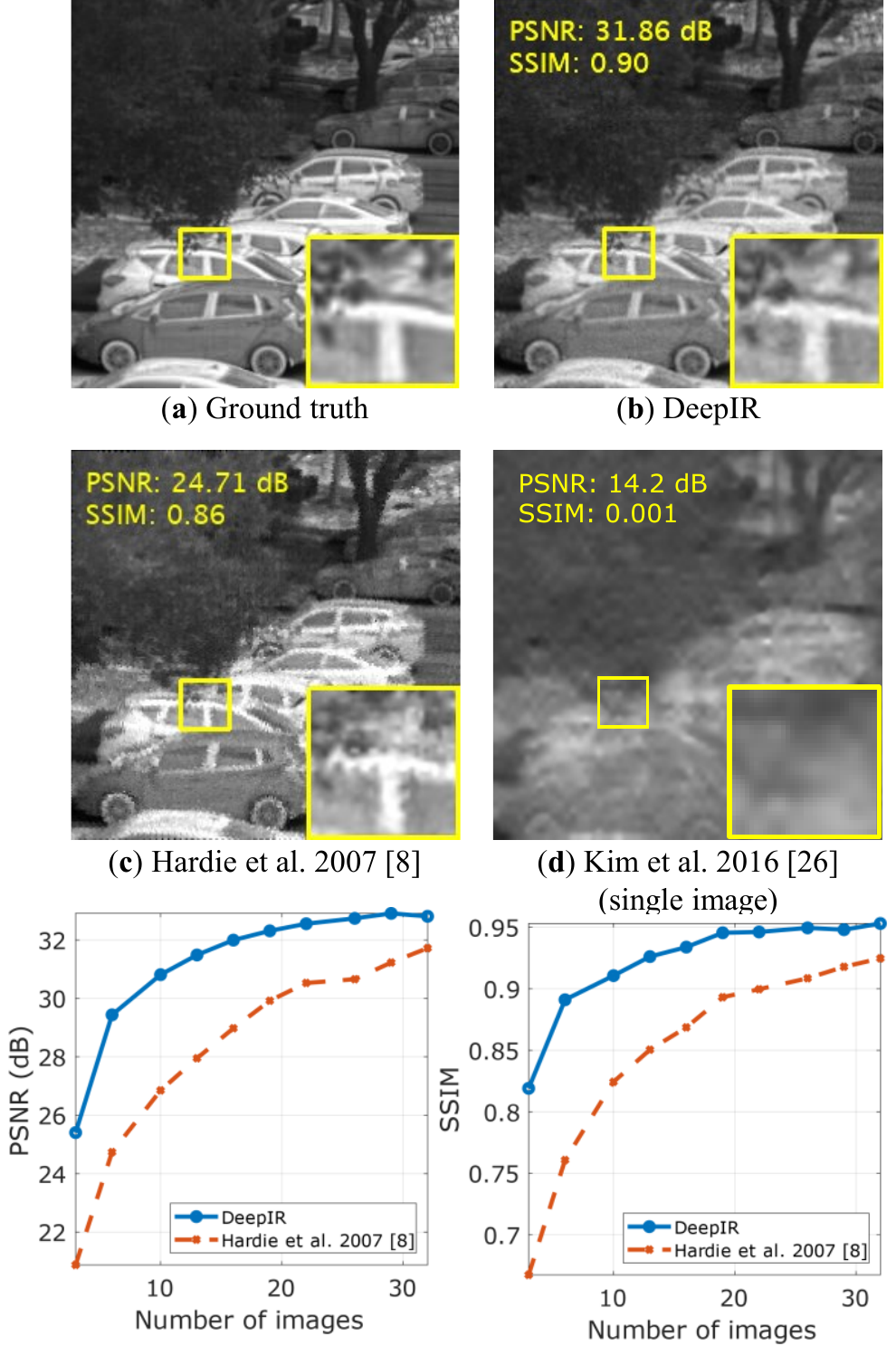}
	\caption{\textbf{Simulations for super resolution.} We simulated a capture of 16 images and performed a $4\times$ super resolution with various approaches. DeepIR outperforms competing techniques including single image super resolution~\cite{kim2016accurate}.}
	\label{fig:sim_sr}
\end{figure}
\bpara{Super Resolution.}
Figure \ref{fig:sim_sr} shows super resolution results on a simulated scene.
We simulated a capture of 16 images, downsampled by $4\times$, and then recovered using various approaches.
We compared against a modified version of Hardie et al.~\cite{hardie2007map} where we added a TV prior to regularize the problem.
We also compared with a single image super resolution (SISR) approach trained on visible images~\cite{kim2016accurate}.
Specifically, we first denoised with He et al.'s \cite{he2018single} approach, and then used this as input to the SISR network.

\subsection{Hardware experiments}
%
We captured images with the FLIR Boson 640 to demonstrate NUC over a wide range of scenes, and with FLIR Lepton 3.5 to demonstrate super resolution.
The Boson camera captured images at $60$ frames per second (fps), and the Lepton captured images at $8.8$ fps.
In all our experiments, we kept the scene static, and only perturbed the camera, creating a jitter in the images.
The Boson camera had an in built mechanical shutter that periodically closes to perform NUC. 
We performed denoising with and without the shutter-based flat field correction, in order to demonstrate the efficacy of DeepIR across varying hardware settings.
\begin{figure}[!tt]
	\centering
	\includegraphics[width=\columnwidth]{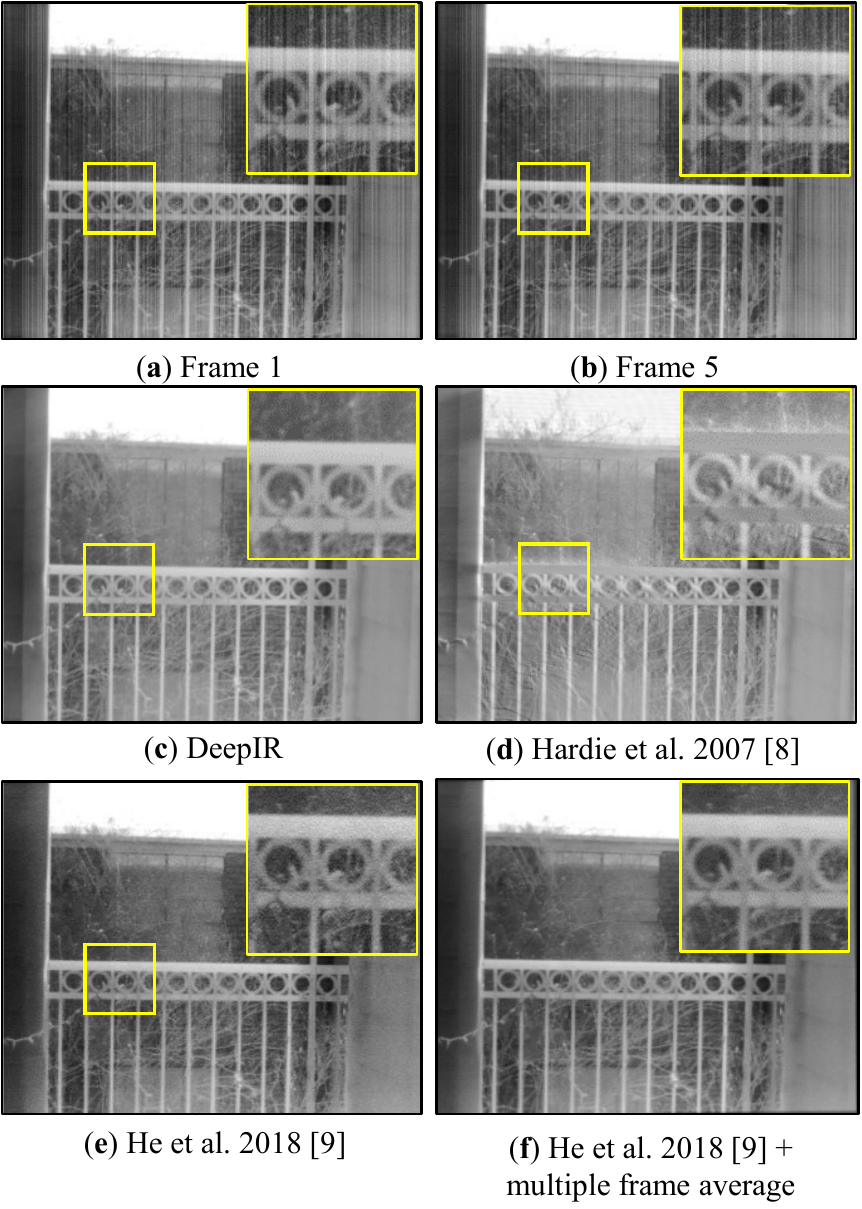}
	\caption{\textbf{Non-uniformity correction without shutter-based compensation.} DeepIR performs comparably to supervised techniques~\cite{he2018single} on cameras without built-in shutter-based FFC.}
	\label{fig:outdoor3}	
\end{figure}
\begin{figure}[!tt]
	\centering
	\includegraphics[width=\columnwidth]{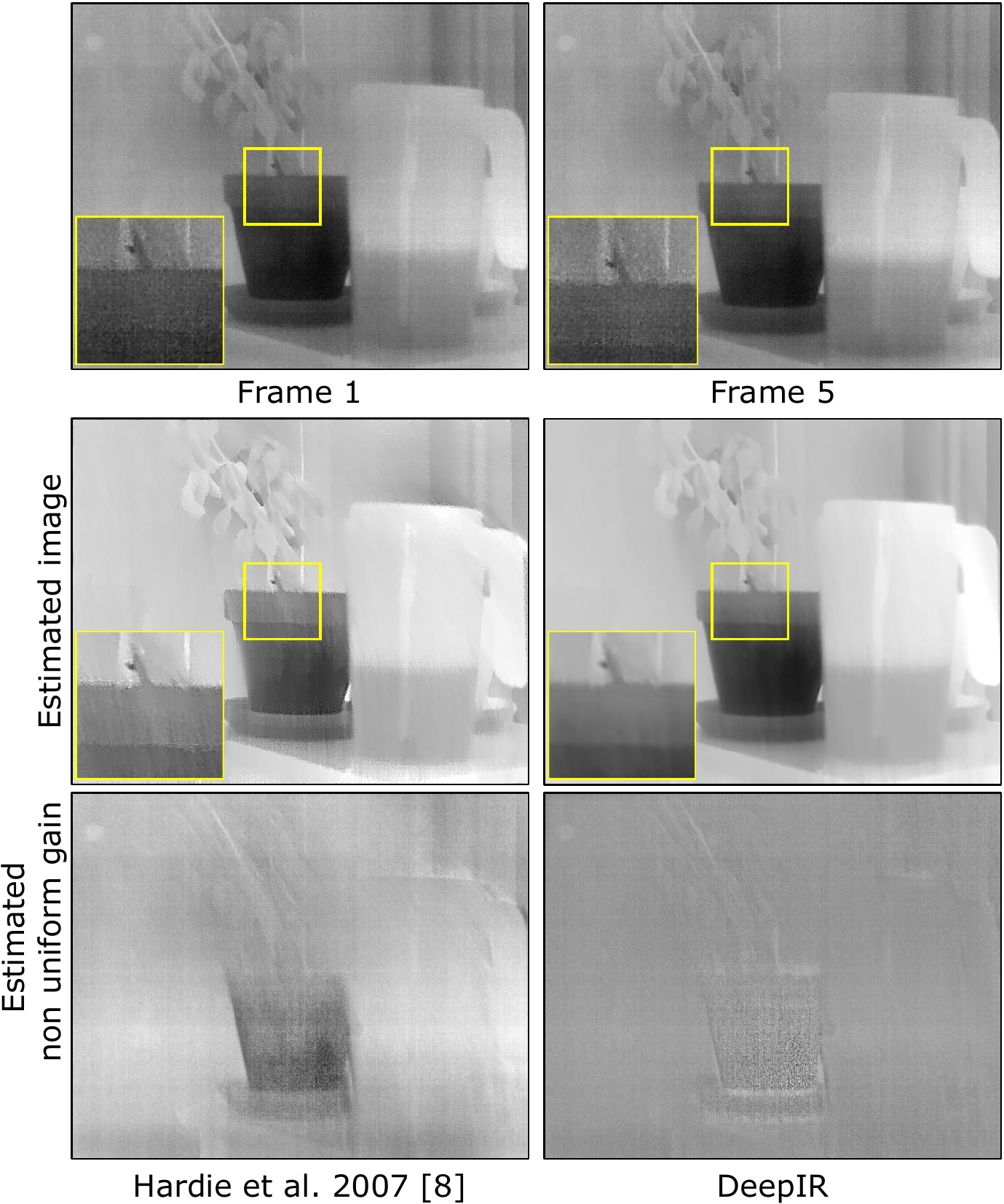}
	\caption{\textbf{Non-uniformity correction with shutter-based compensation.} Even with a shutter-based FFC, scenes with very low temperature variations tend to have residual non uniformities. DeepIR can denoise well even under such conditions.}
	\label{fig:plants1}	
\end{figure}
%

\bpara{Non-uniformity correction.}
Figure \ref{fig:outdoor3} shows the results of non uniformity correction (NUC) with various approaches.
We note that He et al. \cite{he2018single} relied on a large pool of data to learn to specifically remove column-wise NUC.
To keep the comparison fair, we applied this approach to multiple images, registered and averaged them.
We observe that DeepIR outperforms \cite{hardie2007map}, and is comparable to the quality of \cite{he2018single} with multiple frames averaged.

Figure \ref{fig:plants1} shows results with Boson's shutter-based flat field correction applied once during the start of the camera, which largely removed the stripes pattern.
Since the technique in \cite{he2018single} was not trained for such patterns, the approach did not denoise the image.
DeepIR outputs a visibly cleaner image, and estimates a gain that is independent of the scene's geometry.

\bpara{Super resolution.}
The FLIR Lepton camera is a low cost but low resolution thermal imager.
In order to increase the resolution by $4\times$, we employed DeepIR framework with 16 images.
Notice how the low resolution image contains no significant information about the various keys, but the super resolved image has the keys distinctly separated.
DeepIR hence allows us to convert low-cost thermal cameras to high resolution cameras.

\begin{figure}[!tt]
	\centering
	\begin{subfigure}[t]{0.48\columnwidth}
		\centering
		\includegraphics[width=\columnwidth]{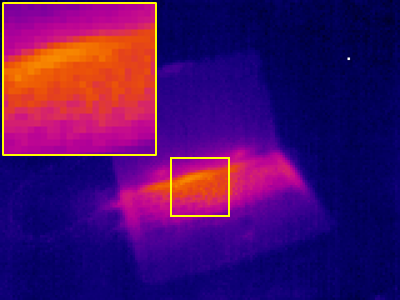}
		\caption{One of the low resolution images.}
	\end{subfigure}
		\begin{subfigure}[t]{0.48\columnwidth}
		\centering
		\includegraphics[width=\columnwidth]{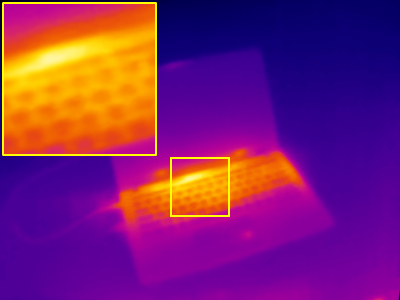}
		\caption{DeepIR super resolved image.  }
	\end{subfigure}
	\caption{\textbf{Super resolution with Lepton camera.} We performed a $4\times$ super resolution with $16$ low resolution images captured with a FLIR Lepton to a resolution of $640\times480$. Images are in false color to show the details. Notice the clearly visible gap in keys.}
	\label{fig:real_sr}
	\vspace{-1.5em}
\end{figure}
\begin{figure}[!tt]
	\centering
	\includegraphics[width=\columnwidth]{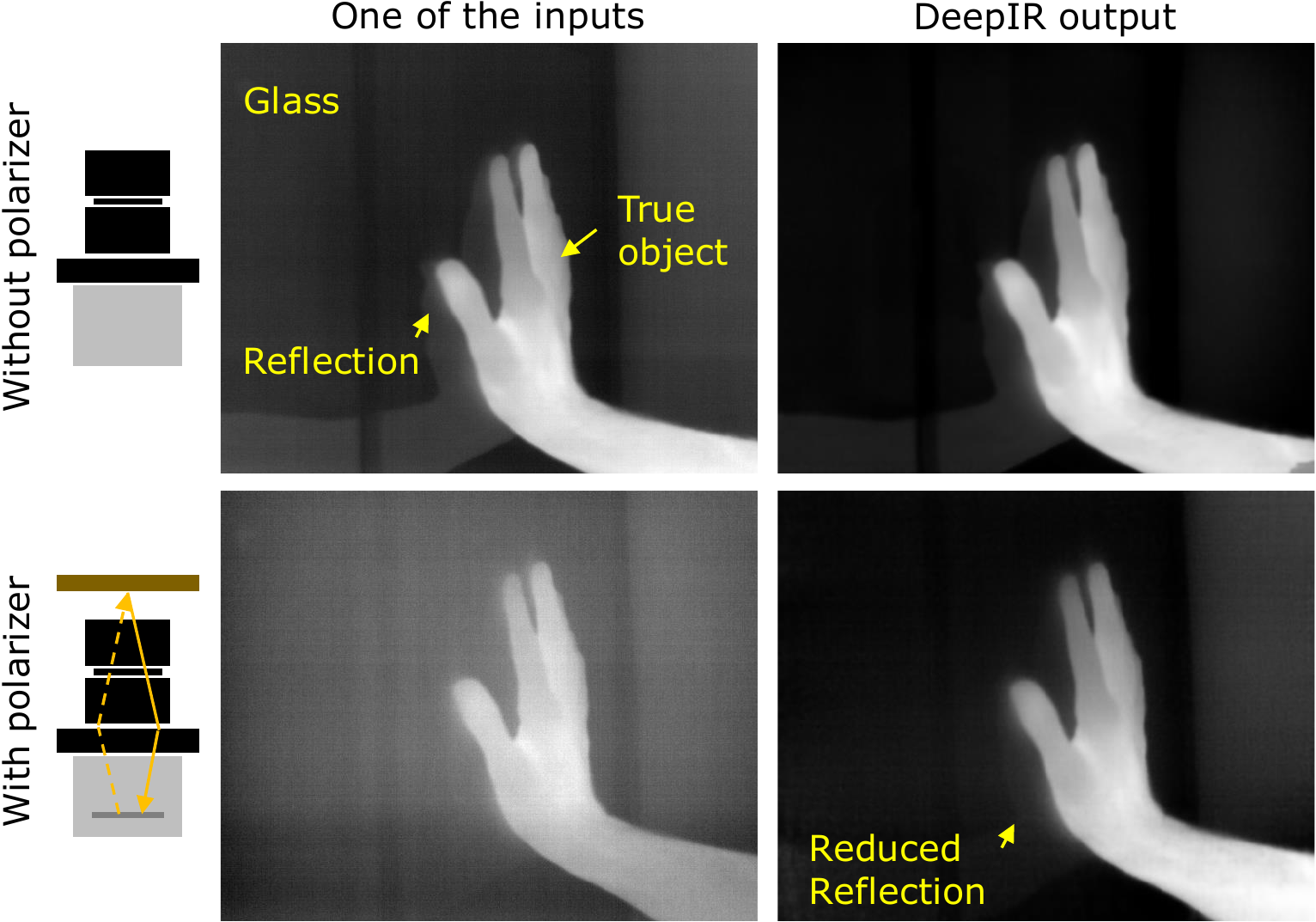}
	\caption{\textbf{NUC with external optics.} Addition of optics such as polarizer causes narcissus effect, which increases the image offset. DeepIR is capable of NUC with quality comparable to imaging without polarizer.}
	\label{fig:polarizer}
	\vspace{-1em}
\end{figure}

\bpara{Suppressing reflections with a polarizer.}
Surfaces such as polished metals and glass are strong reflectors in thermal wavelengths, causing interference.
Since the reflected polarizataion is predominantly orthogonal to the surface, we can utilize a polarizer to remove its effect.
However, the presence of additional optics in front of thermal cameras causes a narcissus effect.
Figure \ref{fig:polarizer} visualizes the effect of a polarizer placed in front of our Boson camera.
This is a compelling example for how the offset term can be highly structured, and hence biased.
We applied DeepIR on the inputs with polarizer to remove the offset term, resulting in an image that was as sharp as the image without a polarizer.

%% file: conclusion.tex
We have developed a general framework called DeepIR for enhancing images in the thermal domain.
We achieved this by noting that camera motion, which is usually a hinderance, can be exploited to our advantage to separate a sequence of images into the scene-dependent radiant flux, and a slowly changing scene-independent non-uniformity.
DeepIR combines the physics of microbolometer sensors, with powerful regularization capabilities by neural network-based representations.
DeepIR relies on the key observation that jittering a camera, while unwanted in visible domain, is highly desirable in the thermal domain as it allows an accurate separation of the sensor-specific non-uniformities from the scene's radiant flux.
We showed compelling results on NUC, super resolution, and correction of narcissus effect with external optics.
Our framework can be applied to several other tasks in thermal imaging including accurate temperature estimation, motion deblurring, and optical flow.

\bpara{Current limitations.} Since our approach relies on neural networks, the optimization process requires several minutes of GPU computations -- resources which preclude a video-rate processing.
Future directions may look into speeding up algorithm with implementation improvements, or by using light-weight networks.

\bpara{Hardware for jittering.} While our approach involved manually jittering the camera, it is possible to build hardware systems that can be jittered internally.
Some examples include an electromagnetic stage that can precisely control the sensor position within the camera, which is already used in cellphone cameras for image stabilization.
Another approach would be to rotate a thick Germanium window in front of the camera that would then produce shifts in various directions.
This was the basis of the jitter camera~\cite{ben2004jitter} and is another promising direction.

%% file: ack.tex
This work was supported by NSF grants CCF-1911094, CCF-1730574, IIS-1838177, IIS-1652633, IIS-1730574, and IIS-2107313; ONR grants N00014-18-12571, N00014-20-1-2534, and MURI N00014-20-1-2787; AFOSR grant FA9550-18-1-0478; and a Vannevar Bush Faculty Fellowship, ONR grant N00014-18-1-2047.

%% file: sup_learning.tex
\begin{figure}[!tt]
	\centering
	\includegraphics[width=\columnwidth]{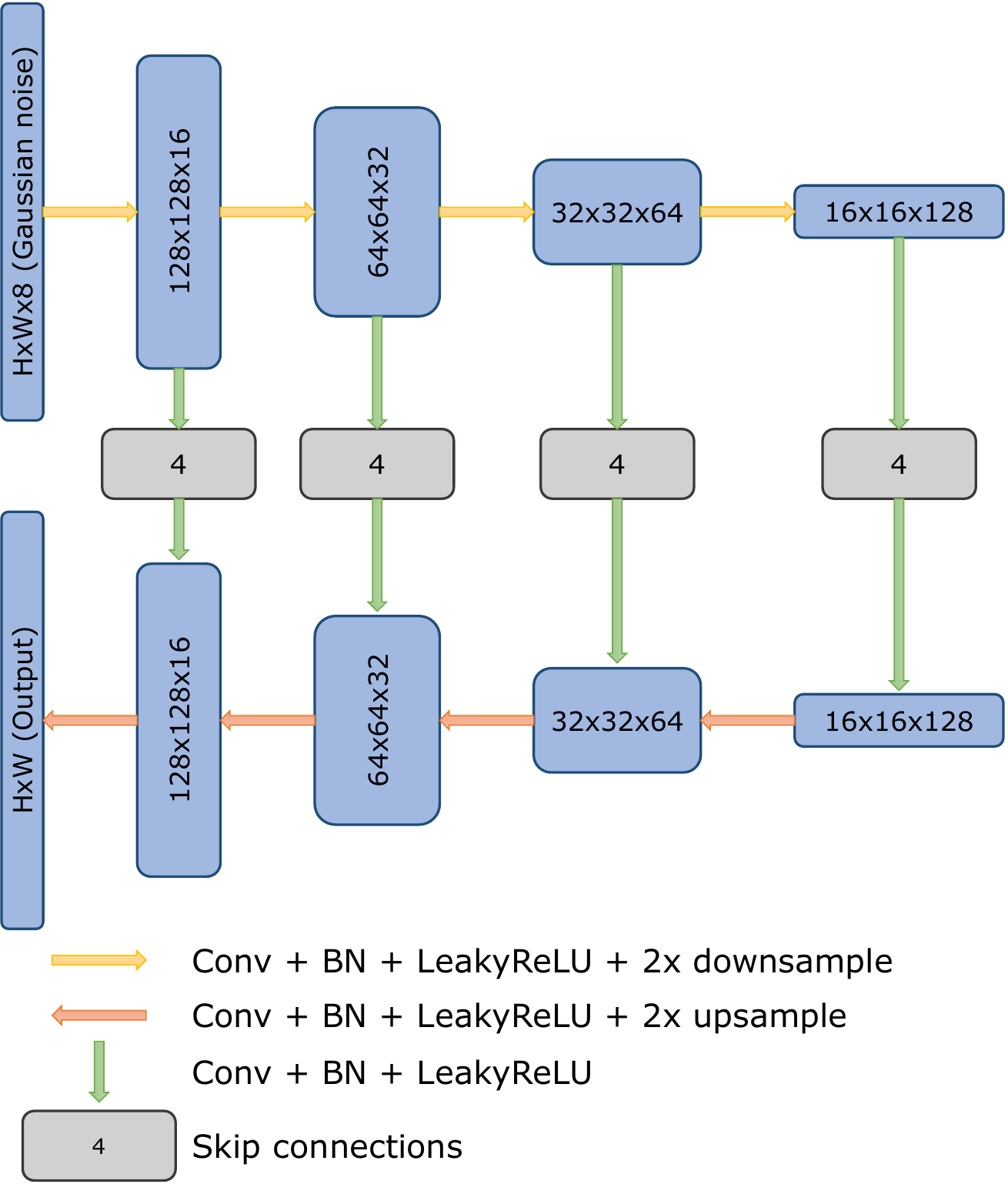}
	\caption{\textbf{Network architecture.} We used the default network architecture proposed in~\cite{ulyanov2018deep} for super resolution.}
	\label{fig:network}
\end{figure}

All the results in the paper were regularized with a deep image prior based regularize.
Our goal was to demonstrate the advantages of combining physics and deep networks, and hence our network architecture was an unmodified version of the architecture utilized in the original paper~\cite{ulyanov2018deep}.
Specifically, we used a convolutional network with skip connections shown in Fig.~\ref{fig:network}.
We note that alternate networks are possibly and potentially capable of giving better results but was not the focus of our paper.

\bpara{Optimization details.}~As mentioned in the paper, we jointly optimized the parameters of the neural network, 6 parameters for each of the $N$ affine matrices, $H\times W$ dimensional gain and offset terms.
The input to the neural network was a $H\times W\times8$ shaped noise that was \textit{not} optimized along with other parameters.
We found that random initialization for affine matrices sufficed -- however to accelerate convergence we first registered the images to the first image using a pyramidal registration algorithm~\cite{thevenaz1998pyramid}.

\bpara{Details about super resolution.}~The image formation model relating the low resolution image $\bfx_k$ and high resolution image $\bfx_\text{HR}$ is,
\begin{align}
	\bfx_k &= \bfg\odot (D M_k \bfx_\text{HR}) + \bfo,
\end{align}
where $D$ is the downsampling operator, and $M_k$ is the transformation matrix.
To prevent aliasing artifacts endemic to downsampling, we chose $D$ as the following operation,
\begin{align}
	X_\text{LR}(u, v) &= \frac{1}{Q^2}\sum_{p=1}^{Q}\sum_{q=1}^{Q} X_\text{HR}(u + p, v + q)
\end{align}
for downsampling by a factor of $Q$.
\begin{figure*}[!tt]
	\centering
	\includegraphics[width=\textwidth]{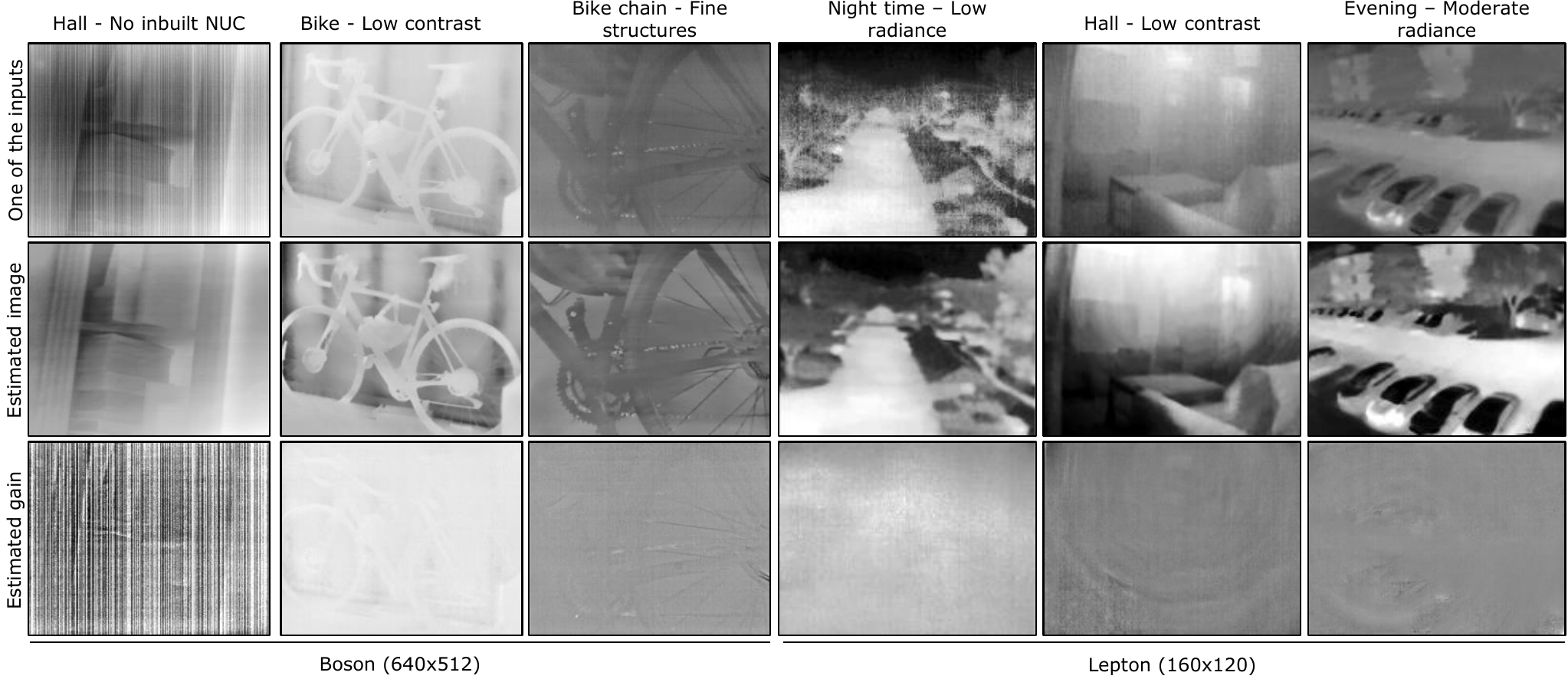}
	\caption{\textbf{NUC on diverse scenes.} Our approach is capable of non-uniformity correction for a wide variety of noise levels and scene complexities.}
	\label{fig:master_nuc}
\end{figure*}
\begin{figure*}[!tt]
	\centering
	\includegraphics[width=\textwidth]{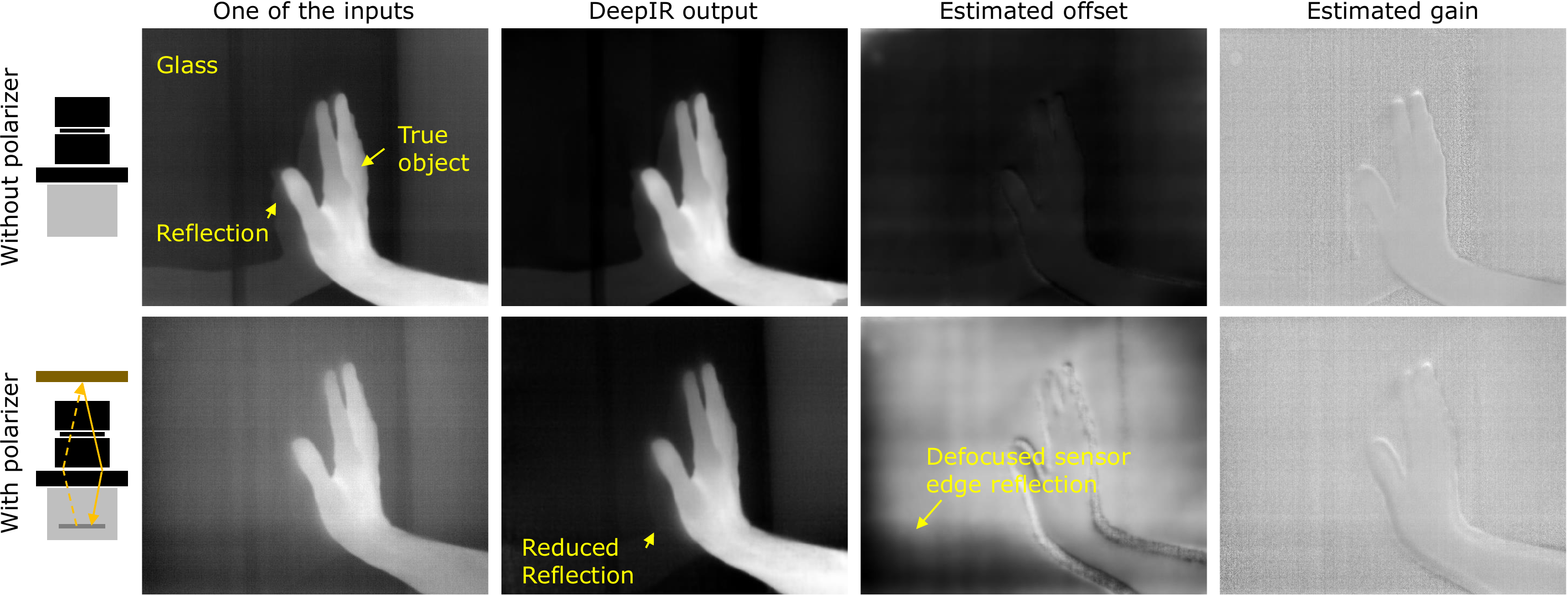}
	\caption{\textbf{Suppressing narcissus effect.} Since we model both gain and offset terms, DeepIR is capable of removing narcissus effects due to external optics like polarizers.}
	\label{fig:master_pol}
\end{figure*}

\bpara{Learning parameters.}~We set the learning rate to $10^{-3}$ and trained for a total of $2,000$ iterations.
For non-uniform correction, there was no penalty for optimizing beyond $2,000$ iterations. 
However increasing the number of iterations proved to be detrimental for super resolution by producing checker-like artifacts in the final reconstruction
This is expected, as deep image prior tends to overfit to noise if run for too many iterations.

Our loss function consisted of MSE loss between predicted image $\bfg \odot M_k\left(\bfx_o + \bfo\right)$ and the ground truth $\bfx_k$, a 2D total variation (TV) prior on the latent image, and a TV loss on the offset term.
The motivation behind the TV loss for the offset is due to it arising from reflections off of optics which tend to be spatially smooth.
We found this to be an effective strategy in separating the gain and offset terms.
We set the weight of the TV loss on the latent image to be $10^{-5}$, and the weight of the TV loss on the offset term to be $10$.
We used a batch size equal to the number of input images.
The model was trained a system with Nvidia RTX 2080 GPU with 8GB memory along with 48GB RAM.
The optimization was implemented with the \verb|pytorch| framework~\cite{NEURIPS2019_9015}.
The code ran for 10 minutes on our computer for five images of size $640\times512$ for a total of $2,000$ iterations.
We will release our optimization code to the public for further research in this direction.

%% file: sup_real.tex
We demonstrate some more results and provide sensitivity to parameters.

\bpara{Hardware details.}~We used the FLIR Boson camera with $640\times512$ spatial resolution capturing images at 60 frames per second (fps), and the FLIR Lepton camera with $160\times120$ spatial resolution capturing images at 9 fps.
We used the \verb|flirpy|~\cite{flirpy} package to control the cameras which allowed us to disable periodic NUC and capture images at full frame rate of the individual cameras.
The Boson camera was equipped with inbuilt flat field correction (FFC), supplementary correction for lens reflections, and temporal noise reduction.
We showed results with and without FFC in the main paper.
In all cases, we disabled temporal noise reduction, as we found that enabling it produced ghosting artifacts.

\bpara{Non-uniformity correction.}
We showed NUC results on some scenes with the Boson camera in the main paper.
We next demonstrate some more experiments to underline the advantages of DeepIR.
Figure~\ref{fig:master_nuc} shows the non-uniformity correction with the various scenes at varying levels of scene complexity.
All experiments included recovery with five images. We found the offset to be nearly zero and hence did not visualize it.
DeepIR performs promisingly in low contrast conditions, absence of inbuilt NUC, low and low radiance levels.


\bpara{Suppressing narcissus effect.}
Figure \ref{fig:master_pol} shows the images with and without polarizer.
Since we model both gain and offset, we were able to suppress the narcissus effect arising out of back reflections from the polarizer.
Notice the defocused edge that is visible in the estimated offset in the image captured with a polarizer.
The edge artifacts looking like the hard were due to minor motion between frames, and can be corrected with a more accurate model of transformation such as optical flow.